\documentclass[%
 reprint,
superscriptaddress,
longbibliography,
 amsmath,amssymb,
prb,
]{revtex4-2}

\usepackage{graphicx}
\usepackage{dcolumn}
\usepackage{bm}
\usepackage{hyperref}
\hypersetup{
	colorlinks=true,
	linkcolor=blue,
	citecolor=blue,
	filecolor=blue,      
	urlcolor=blue,
}
\usepackage{hypcap}
\usepackage{mathrsfs} 
\newcommand{\commented}[1]{}
\usepackage[dvipsnames]{xcolor} 
\newcommand{\note}[1]{#1}

\begin{document}

\preprint{APS/123-QED}


%
%
\title{Dynamical theory of angle-resolved electron energy loss and gain spectroscopies of phonons and magnons \note{in transmission electron microscopy} including multiple scattering effects}


%
%
\author{Jos\'e \'Angel Castellanos-Reyes}
\email{angel.castellanos.research@gmail.com}
\affiliation{Department of Physics and Astronomy, Uppsala University, Box 516, 75120 Uppsala, Sweden}

\author{Paul Zeiger}%
\email{paul.zeiger@physics.uu.se}
\affiliation{Department of Physics and Astronomy, Uppsala University, Box 516, 75120 Uppsala, Sweden}

\author{J\'an Rusz}
\email{jan.rusz@physics.uu.se}
\affiliation{Department of Physics and Astronomy, Uppsala University, Box 516, 75120 Uppsala, Sweden}

\date{\today}
%
%
\begin{abstract}
We present a method for computing angle-resolved electron energy loss and gain spectroscopies for phonon and magnon excitations in transmission electron microscopy. Fractional scattering intensities are derived from the temperature-dependent time auto-correlation of the electron beam wave function. This method captures both single and multiple scattering processes, as well as dynamical diffraction effects. Our method remains computationally efficient, and it is easy to parallelize.

\end{abstract}
\maketitle

Recent instrumental advances in scanning transmission electron microscopy (\note{STEM};~\cite{krivanek2014nature,plotkin-swing_hybrid_2020}) have enabled the study of low-energy excitations, including phonons. Electron energy loss and gain spectroscopies (EELS and EEGS, respectively) frequently emerge as preferred methods, especially when high spatial resolution is a priority. \note{While reflection electron energy loss spectroscopy (REELS) is effective at surfaces and in high-resolution contexts~\cite{ref1,ref2}, it is limited by the low penetration depth of low-energy electrons, typically not exceeding 10~eV, which restricts it to surface or ultrathin film studies. This limitation is not present in STEM techniques}. Atomic resolution vibrational \note{STEM-EELS} of single defects or impurities, sensitivity to isotopic composition down to the nano-scale and nano-scale phonon dynamics are among the breakthroughs enabled by the new technology~\cite{yan_single-defect_2021,hage_single-atom_2020,hachtel_identification_2019,senga_imaging_2022,gadre_nanoscale_2022}. \note{Magnons, which play a crucial role in the emerging field of magnonics \cite{Barman2021}, represent another type of low-energy excitation that holds promise for measurement with this technique in the near future \cite{lyon_theory_2021, castellanos_unveiling_2023}.}

Theoretical frameworks are crucial for modeling and understanding the observed low-energy spectra, providing a foundational basis for the interpretation of experiments. However, existing computational approaches to electron scattering often rely on restrictive approximations, making quantitative comparisons with experiments challenging. These approaches are frequently confined to first-order perturbative treatments or semiclassical limits. Moreover, they frequently lack one or more important aspects of electron scattering, such as energy resolution, multiple phonon scattering, multi-phonon scattering~\note{\footnote{\note{With \textit{multiphonon scattering}, we refer to the situation in which the electron (from the electron beam) scatters creating several phonons in a single inelastic event. \textit{Multiple phonon scattering} corresponds to multiple inelastic scattering events, in which the electron creates one or more phonons each time.}}}, dynamical diffraction, and momentum-transfer selectivity~\cite{nicholls_theory_2019,senga_position_2019,hage_single-atom_2020, allen_inelastic_1995,martin_model_2009,forbes_modeling_2016,dwyerProspectsSpatialResolution2017,rez_lattice_2021,forbes_quantum_2010,lugg_atomic_2015,zeiger_efficient_2020,zeiger_frequency_2021,lubk,zeiger2023lessons}.

In this Letter, we show that via the time auto-correlation of the electron beam wave function, it is possible to describe angle-resolved phonon energy loss and energy gain processes \note{in transmission electron microscopy} including processes involving multiple phonons, effects of temperature and dynamical diffraction. Furthermore, we show how to extend the method to simulate angle-resolved magnon EELS and EEGS. Moreover, by its nature, this approach is computationally efficient and easy to deploy on parallel computers.

We first build the method within the framework of the quantum excitation of phonons (QEP) model \cite{forbes_quantum_2010,lugg_atomic_2015}. \note{As such, our method described below inherits all the flexibility as well as approximations assumed in the QEP model.} The key step of QEP is an approximation akin to Born-Oppenheimer theory. 
The method then introduces the concept of configurations, denoted as $\bm{\tau}$, which are composite vectors comprising the positions of all nuclei in the system. It also involves an auxiliary electron-beam wave function, denoted as $\phi(\mathbf{r},\bm{\tau})$, or its position-to-momentum Fourier transform $\phi(\mathbf{q},\bm{\tau})$, which parametrically depends on the configuration $\bm{\tau}$. For any given arbitrary configuration $\bm{\tau}$, the auxiliary electron beam wave function can be calculated, for example, by a standard multislice method \cite{forbes_quantum_2010,lugg_atomic_2015,cowley_scattering_1957}. 

Although the QEP model operates with fixed positions of nuclei, it provides a rigorous quantum mechanical treatment considering all possible crystal states $|\mathbf{n}\rangle$, where $\mathbf{n}$ is a composite vector describing the state of the crystal in Fock space. Thus, $\mathbf{n}$ consists of non-negative integers, expressing the number of excited quanta of all available vibrational modes. The amplitude of probability for a configuration $\bm{\tau}$ to be realized in the crystal state $|\mathbf{n}\rangle$ of energy $E_\mathbf{n}$ is expressed through the crystal wave functions $a_\mathbf{n}(\bm{\tau}) = \langle \bm{\tau} | \mathbf{n} \rangle$, which can be written as a product of real-valued Hermite polynomials~\cite{lugg_atomic_2015}. Within this formalism, the QEP model provides the following expression for the inelastically scattered wave corresponding to a transition of the crystal from state $|\mathbf{n}\rangle$ to state $|\mathbf{m}\rangle$:
%
%
%
\begin{equation}
\psi_{\mathbf{n}\mathbf{m}}(\mathbf{q}) = \int
a_{\mathbf{m}}^*(\bm{\tau})
\phi(\mathbf{q},\bm{\tau})
a_{\mathbf{n}}(\bm{\tau})
d\bm{\tau},
\label{eq:psimn}
\end{equation}
%
see for example Eq.~(15) in Ref.~\cite{lugg_atomic_2015}. By introducing the \emph{beam transmission operator} $\hat{\phi}(\mathbf{q})$, defined by
$\langle \bm{\tau} | \hat{\phi}(\mathbf{q}) | \bm{\tau'} \rangle = \phi\left(\mathbf{q},\bm{\tau}\right) \delta(\bm{\tau-\tau'}),$ Eq.~\ref{eq:psimn} can be concisely rewritten using Dirac notation as
%
%
%
$\psi_{\mathbf{n}\mathbf{m}}(\mathbf{q})=\langle \mathbf{m} | \hat{\phi}(\mathbf{q}) | \mathbf{n} \rangle.$

Within this notation, the total scattering cross-section is given in the QEP model as \cite{lugg_atomic_2015}
%
%
%
\begin{eqnarray}
  I_\text{tot}(\mathbf{q}) & = & \frac{1}{Z} \sum_\mathbf{n,m} e^{-\beta E_\mathbf{n}} |\langle \mathbf{m} | \hat{\phi}(\mathbf{q}) | \mathbf{n} \rangle|^2 
  \label{eq:sumint} \\
  & = & \frac{1}{Z} \sum_\mathbf{n} e^{-\beta E_\mathbf{n}} \langle \mathbf{n} | \hat{\phi}^\dagger(\mathbf{q}) \hat{\phi}(\mathbf{q}) | \mathbf{n} \rangle \nonumber \\
  & = & \int d\bm{\tau} \left[ \frac{1}{Z} \sum_\mathbf{n} e^{-\beta E_\mathbf{n}} |a_\mathbf{n}(\bm{\tau})|^2 \right] |\phi(\mathbf{q},\bm{\tau})|^2, 
\label{eq:Itot}
\end{eqnarray}
%
where $Z=\sum_\mathbf{n} e^{-\beta E_\mathbf{n}}$, $\beta = (k_B T)^{-1}$, $k_B $ is the Boltzmann constant, and the expression in square brackets represents the thermally averaged probability $P_T(\bm{\tau})$ that the system is in the configuration $\bm{\tau}$ at temperature $T$. The expression \ref{eq:Itot} is formally equivalent to the frozen phonon model~\cite{loane_thermal_1991}, where $|\phi(\mathbf{q},\bm{\tau})|^2$ is averaged over random uncorrelated samples from the configuration space, respecting the probability distribution $P_T(\bm{\tau})$.

The given expressions for $I_\text{tot}$ yield the total scattered intensity, encompassing the sum over all possible elastic and inelastic transitions. However, they do not provide spectroscopic information.  Moreover,  Eq.~\ref{eq:psimn} expresses the inelastic wave in a manner that prevents introducing any form of probability for the configuration $\bm{\tau}$, since the product $a_{\mathbf{m}}^*(\bm{\tau})a_{\mathbf{n}}(\bm{\tau})$ is a real number that can be negative.

Nevertheless, it is possible to extract the spectroscopic information starting from Eq.~\ref{eq:sumint}: Since the transition from state $|\mathbf{n}\rangle$ to state $|\mathbf{m}\rangle$ corresponds to an energy loss (or gain) $E = E_\mathbf{m}-E_\mathbf{n}$, the transitions can be formally sorted on the energy axis by including~$\delta(E_\mathbf{m}-E_\mathbf{n}-E)$ to associate individual crystal state transitions with a specific energy transfer $E$. Therefore, 
%
%
\begin{equation}
I_\text{tot}(\mathbf{q}) = \int I(\mathbf{q},E) dE,
\label{eq.ourItot}
\end{equation}
%
%
where
%
\begin{eqnarray}
  &\phantom{\!}&I(\mathbf{q},E)  =  \frac{1}{Z} \sum_\mathbf{n,m} e^{-\beta E_\mathbf{n}} |\langle \mathbf{m} | \hat{\phi}(\mathbf{q}) | \mathbf{n} \rangle|^2 \delta(E_\mathbf{m}-E_\mathbf{n}-E) \label{eq:specsum} \nonumber \\
  & = & \sum_\mathbf{n,m} \frac{e^{-\beta E_\mathbf{n}}}{Z} \int_{-\infty}^{\infty} \frac{e^{\frac{i}{\hbar}(E_\mathbf{m}-E_\mathbf{n}-E)t}}{2\pi\hbar} |\langle \mathbf{m} | \hat{\phi}(\mathbf{q}) | \mathbf{n} \rangle|^2 dt \nonumber \\
  & = & \sum_\mathbf{n} \frac{e^{-\beta E_\mathbf{n}}}{Z} \int_{-\infty}^{\infty} \frac{e^{-\frac{i}{\hbar}Et}}{2\pi\hbar} \langle \mathbf{n} | \hat{\phi}^\dagger(\mathbf{q}) \hat{U}_c^\dagger(t) \hat{\phi}(\mathbf{q}) \hat{U}_c(t) | \mathbf{n} \rangle dt 
  \nonumber \\
  & = & \sum_\mathbf{n} \frac{e^{-\beta E_\mathbf{n}}}{Z} \int_{-\infty}^{\infty} \frac{e^{-\frac{i}{\hbar}Et}}{2\pi\hbar} \langle \mathbf{n} | \hat{\phi}^\dagger(\mathbf{q},0) \hat{\phi}(\mathbf{q},t) | \mathbf{n}  \rangle dt 
  \nonumber \\
  & = & \int_{-\infty}^{\infty} \frac{e^{-\frac{i}{\hbar}Et}}{2\pi\hbar} \frac{1}{Z} \mathrm{Tr} \left[ e^{-\beta \hat{H}_c} \hat{\phi}^\dagger(\mathbf{q},0) \hat{\phi}(\mathbf{q},t) \right] dt
  \label{eq:ensemble}
   \\
  & = & \int_{-\infty}^{\infty} \frac{e^{-\frac{i}{\hbar}Et}}{2\pi\hbar}
  c_{\phi\phi}(t)
  dt
   \nonumber \\
  & = & 
  c_{\phi\phi}(E)
  , \label{eq:ensemble2}
\end{eqnarray}
%
in which $\hat{U}_c(t)=e^{-\frac{i}{\hbar}\hat{H}_c t}$ is the crystal time-evolution operator, $c_{\phi\phi}(t)=\frac{1}{Z} \mathrm{Tr} \left[ e^{-\beta \hat{H}_c} \hat{\phi}^\dagger(\mathbf{q},0) \hat{\phi}(\mathbf{q},t) \right]$ is the quantum-mechanical time auto-correlation function of the operator $\hat{\phi}(\mathbf{q})$ in Heisenberg representation, and $c_{\phi\phi}(E)$ the time-to-energy Fourier transform of $c_{\phi\phi}(t)$. 

Correlation functions that depend on the dynamics of nuclei \cite{vanhove_correlations_1954} are often approximated using molecular dynamics (MD) methods or, alternatively, by ring-polymer molecular dynamics (RPMD), which incorporates zero-point-energy effects of nuclei \cite{craig2004}. These methods yield a Kubo-transformed correlation function~\cite{kubo_statistical_1957}, expressed in our case as:
%
%
%
\begin{equation} 
 \tilde{c}_{\phi\phi}(t) = \frac{1}{\beta Z} \int_0^\beta \mathrm{Tr} \left[ e^{-(\beta-\lambda) \hat{H}_c} \hat{\phi}^\dagger(\mathbf{q},0) e^{-\lambda \hat{H}_c} \hat{\phi}(\mathbf{q},t)  \right] d\lambda
 .
 \label{eq:kubo}
\end{equation}
%
The following exact relation holds between the time-to-energy Fourier-transforms of the two aforementioned correlation functions~\cite{berens_molecular_1981,craig2004}
%
%
%
\begin{equation} 
  c_{\phi\phi}(E) = \frac{\beta E}{1-e^{-\beta E}} \tilde{c}_{\phi\phi}(E) 
  ,
  \label{eq:kubo_vs_normal}
\end{equation}
%
which allows the recovery of the initial correlation function appearing in Eq.~\ref{eq:ensemble2}.

The configurational average in the correlation function is approximated through a time average over a classical trajectory. 
%
Employing the convolution theorem, $\tilde{c}_{\phi\phi}(E)$ can then be obtained as the squared amplitude of the time-to-energy Fourier transform of the time-dependent multislice wave function $\phi_\text{MD}(\mathbf{q},t)$, whose time-dependence is realized via its parametric dependence on the position of atoms, which naturally depend on time in a MD simulation. Therefore,
%
%
%
\begin{equation}
    I(\mathbf{q},E) \propto \frac{\beta E}{1-e^{-\beta E}} | \phi_\text{MD}(\mathbf{q},E) |^2.
    \label{eq:theeq}
\end{equation}
%
While this approximation is exact for harmonic potentials and operators linear in position or momentum, errors in $\tilde{c}_{\phi\phi}(t)$ due to neglected quantum phases become more significant for larger times $t$ beyond these conditions~\cite{miller_semiclassical_2001,craig2004}. Typically, a thermal time $\beta\hbar$ serves as a benchmark, beyond which the correlation functions calculated by MD methods start losing their accuracy \cite{craig2004}.


The derivation and discussion above lead to the following procedure to simulate spectra: In the first step, a sufficiently long MD trajectory of a structure's supercell is calculated using empirical, machine learning, or even \emph{ab-initio} interatomic forces. The length of the MD trajectory $t_\text{tot}$ determines the lowest nonzero vibrational energy $E_\text{min}=2\pi\hbar/t_\text{tot}$ that can be obtained via the discrete Fourier transform, which matches then the frequency resolution of the calculated spectra. In contrast to frozen phonon methods~\cite{loane_thermal_1991,muller_simulation_2001}, where the atomic displacements $\bm{\tau}_i$ in structure snapshots should be uncorrelated, here the snapshots are taken at short time intervals $\Delta t \sim 10^{-14}$~s to ensure that the maximum energy $E_\text{max}=\pi\hbar/\Delta t$ in the discrete Fourier transform is greater than the highest energies in the phonon density of states. Therefore, subsequent snapshots have in general strongly correlated atomic displacements.

In the second step, the electron beam is propagated through individual snapshots, for example, using a multislice method~\cite{cowley_scattering_1957,barthel_dr_2018}. Accumulating a set of exit wave functions for all snapshots results in a three-dimensional dataset $\phi_\text{MD}(q_x,q_y,t)$, which is Fourier transformed from the time to the energy domain, producing $\phi_\text{MD}(q_x,q_y,E)$. The squared amplitude of this object gives the approximate $\tilde{c}_{\phi\phi}(E)$, which is then converted to $c_{\phi\phi}(E)$ via Eq.~\ref{eq:kubo_vs_normal}, yielding $I(\mathbf{q},E)$ of Eq.~\ref{eq:ensemble2} up to a normalization factor. A convenient choice is to work with a fractional intensity, where
%
%
%
$$
\iint I(\mathbf{q},E)  d\mathbf{q} dE \stackrel{!}{=} 1,
$$
%
which is then imposed on the $c_{\phi\phi}(E)$ evaluated on a finite numerical grid of momentum and energy transfers. 
The integral over energies covers both negative (energy gains) and positive values (energy losses).

To improve the statistics, it is typically necessary to average $c_{\phi\phi}(E)$ over several MD trajectories. Various strategies can be employed for this purpose. One approach involves simulating a canonical ensemble, from which several uncorrelated snapshots of the phase space (positions and momenta) are selected. These snapshots would then serve as starting points for separate micro-canonical MD calculations. The advantage of this strategy is that there are no forces acting on the atoms other than those dictated by the interatomic potential. An alternative approach consists of simulating a single longer canonical MD trajectory using Langevin dynamics with low damping. Segments of this trajectory can then be used to evaluate individual $I(\mathbf{q},E)$ for subsequent averaging. Mild damping minimally affects the system dynamics, typically blurring the phonon density of states~\cite{zeiger_frequency_2021}, while directly sampling the canonical ensemble assumed in Eq.~\ref{eq:ensemble}.

We exhibit an implementation of this method by simulating \note{thermal diffuse scattering for a silicon crystal, as well as} electron energy loss and gain spectra arising from phonon excitations in widely studied \note{hexagonal boron nitride (hBN)}. We show that it provides full momentum- and energy-resolved $I(\mathbf{q},E)$ information, including the elastic channel, as well as the energy loss and gain channels, consistent with the results of \note{the} frequency-resolved frozen phonon multislice method (FRFPMS; \cite{zeiger_efficient_2020}). Furthermore, the energy-integral of $c_{\phi\phi}(E)$ matches closely the results of the standard QEP method applied to snapshots of \note{the} correlated motion of atoms.

MD simulations \note{for silicon} \cite{LAMMPS_paper_2022} \note{were} performed on a $7 \times 7 \times \note{92}$ supercell of cubic unit cells of silicon, with \note{the $x$, $y$ and $z$ directions of the cell aligned with the [100], [010], and [001]-directions, respectively, and} a lattice parameter of 5.4773~\AA{}, which was obtained as an average lattice parameter in a $NPT$ simulation at temperature $T=300$~K and zero pressure. 
%
%
\footnote{\note{The machine learning SNAP interatomic potential~\cite{thompson_interatomic_2015} for silicon from Ref.~\cite{zuo_performance_2020} was employed. A time step of 0.5~fs was chosen, and after 5000 steps of thermalization, a 0.1~ns trajectory at 300K was generated to capture snapshots. A relatively low Langevin damping of 0.5~ps was used to minimize its impact on the dynamics. In total, 4000 snapshots were generated with a $\Delta t=25$~fs, allowing to reach vibrational frequencies of up to 20~THz ($\sim 83$~meV). For all of them, we have calculated the exit wave functions using \textsc{DrProbe}~\cite{barthel_dr_2018} on a numerical grid of $560 \times 560$ using \note{$736$} slices across the whole thickness of the supercell. To evaluate $I(\mathbf{q},E)$, we have averaged over 79 sets of 100 consecutive snapshots (i.e., $T=2.5$~ps), mutually offset by 50 snapshots. In this way, all the 4000 snapshots were utilized twice, except for the first and the last 50 snapshots. The same MD trajectory was used for FRFPMS calculations, performed on a grid of 1~THz wide frequency bins covering the range of up to 18~THz. Within each frequency bin, 128 snapshots have been generated by band-pass filtering the MD trajectory.}
}

%
%
%
\begin{figure}
    \centering
    \includegraphics[width=\columnwidth]{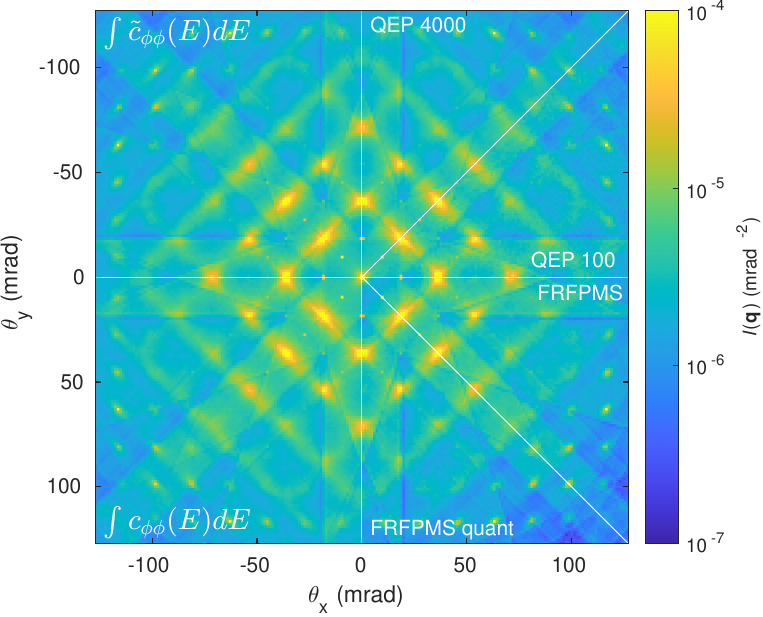}
    \includegraphics[width=\columnwidth]{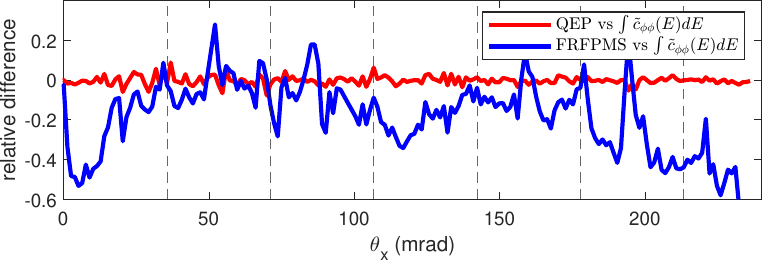}
    \caption{Top panel: Diffraction pattern of a \note{50}~nm thick silicon crystal in $[001]$ zone axis. Individual segments show an energy integral of $\tilde{c}_{\phi\phi}(E)$ and $c_{\phi\phi}(E)$, results of the standard QEP method using 100 evenly spaced snapshots taken every 1~ps (QEP 100) or using 4000 snapshots taken every 25~fs (QEP 4000), and finally the results of an energy integral of FRFPMS method as well as FRFPMS with quantum statistics corrections (FRFPMS quant; see Eq.~53 in Ref.~\cite{zeiger2023lessons}). Bottom panel: \note{relative difference (difference divided by sum)} of line profiles extracted from models indicated in the legend along the $\theta_y=0$ line. \note{Vertical dashed lines indicate positions of Bragg spots $(2k,0,0)$.}}
    \label{fig:difpat}
\end{figure}
%

\note{Similarly, MD simulations were performed on a 8$\times$14$\times$30 supercell of the orthogonal unit cell of hBN in AA' stacking with lattice parameters $a=1.447$~\AA{} and $c=6.736$~\AA, which were determined in a $NPT$ ensemble run on a 8$\times$14$\times$6 supercell, similar to silicon.}
%
%
\note{
\footnote{
\note{The $x$, $y$, and $z$-dimensions were aligned with the $[2\bar{1}\bar{1}0]$, $[01\bar{1}0]$ and $[0001]$-directions, respectively. The interatomic potential was a machine-learned, so-called GAP potential for hBN \cite{bartokGaussianApproximationPotentials2010,bartokRepresentingChemicalEnvironments2013,thiemannMachineLearningPotential2020} and the time step was set to 0.5~fs in all simulations. For the results shown here, one trajectory of 100~ps was simulated after an initial thermalization of 10~ps employing a Langevin thermostat with a damping of 0.5~ps. The positions of all atoms were saved every 7.5~fs, which allows to reach vibrational frequencies up to 66.67~THz ($\sim 276$~meV) for a total of 13332 snapshots. As in the case of silicon, we calculated the beam exit wave function for all snapshots using \textsc{DrProbe} on a numerical grid of $560 \times 560$, albeit using only $500$ slices across the whole thickness of the supercell. $I(\mathbf{q},E)$ was evaluated as an average over 79 sets of 332 consecutive snapshots (i.e., $T=2.49$~ps), mutually offset by 116 snapshots. Snapshots for FRFPMS simulations were sampled using band-pass filtering for 51 energy bins between 0 and 50~THz with a width of 1~THz per bin. The 0~THz bin considers thereby only frequencies between 0.0 to 0.5~THz.}}}

Figure~\ref{fig:difpat} shows the diffraction pattern \note{for our silicon model} illuminated by a 60~kV plane-wave electron beam obtained from Eq.~\ref{eq.ourItot}, FRFPMS, and QEP calculations. The most striking difference concerns the FRFPMS method, predicting intensities that decline with increasing scattering angle faster than in all other methods. FRFPMS has been shown to provide an accurate description of single-phonon excitations, while multi-phonon processes are to large extent missing \cite{zeiger2023lessons}. The close numerical match between Eq.~\ref{eq.ourItot} and QEP methods suggests that the intensity missing in FRFPMS is due to such multi-phonon processes. The more detailed view offered by \note{relative differences shown in the} bottom panel of Fig.~\ref{fig:difpat} reveals that FRFPMS underestimates also the inelastic intensity in the closest vicinity of the direct beam. 
%
%
We defer a more detailed discussion of the observed differences to future research.
\begin{figure}
    \centering
    \includegraphics[width=\columnwidth]{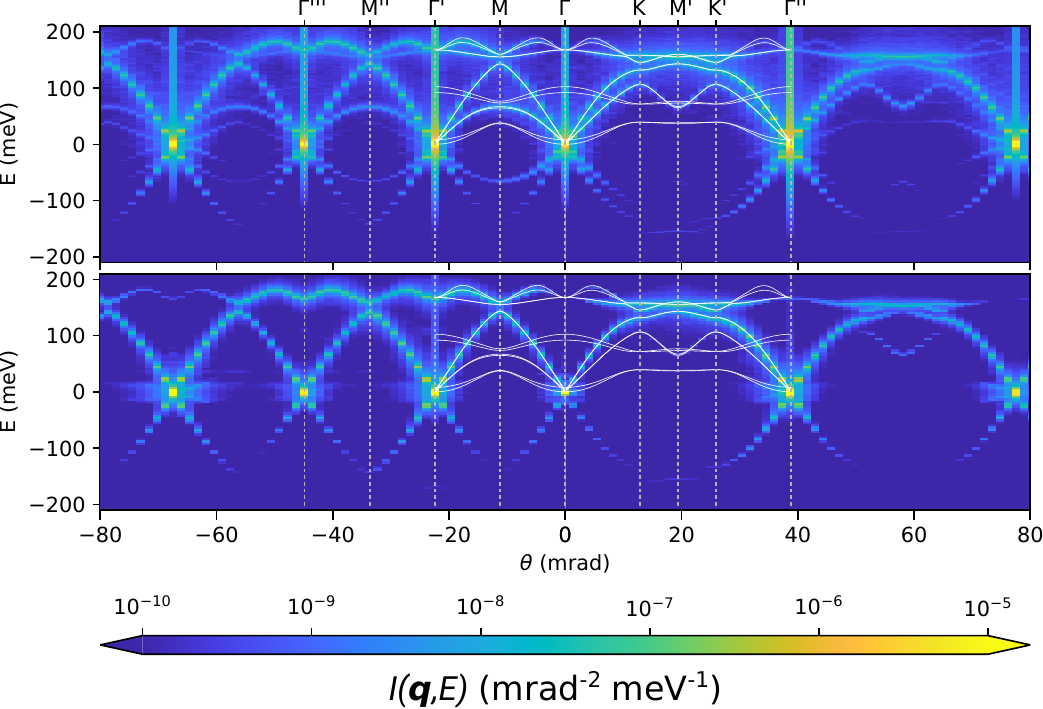}
    \caption{
    \note{
    (top) Phonon dispersion $I(\mathbf{q}, E)$ of a 20~nm thick hBN in [0001] zone axis along the two distinct directions $\Gamma$--M--$\Gamma'$ and $\Gamma$--K--M'--K'--$\Gamma''$ (white curves)
    computed according to Eqs.~\eqref{eq:ensemble2}-\eqref{eq:theeq}. (bottom) The corresponding FRFPMS calculation, including quantum corrections as proposed in Ref.~\cite{zeiger2023lessons}. 
    }
    }
    \label{fig:qEplot}
\end{figure}
%

\note{In order to} highlight the energy-loss and -gain information provided by our method, we show in \note{the top panel of} Fig.~\ref{fig:qEplot} a visualization of the scattering cross-section \note{for our hBN model structure} on selected planes within the $(q_x,q_y,E)$ space, and compare it with corresponding results of FRFPMS calculations \note{(bottom panel)}. 
In both cases, the intensities show local maxima following the characteristic features of the phonon band dispersion of bulk \note{hBN \footnote{We have computed the phonon dispersion for the GAP potential using phonopy with its LAMMPS interface \cite{togoImplementationStrategiesPhonopy2023,togoFirstprinciplesPhononCalculations2023}.}}. \note{The match between both methods} is very close \note{for LA and LO modes}, except for the higher energy resolution provided by $c_{\phi\phi}(E)$ and the appearance of background intensity, which we ascribe to processes involving excitations of multiple phonons, in analogy with \note{the} analysis of inelastic neutron scattering \cite{dawidowski_efficient_1998}. 
\note{Notably, unlike the FRFPMS results, the top panel aligns with experimental observations~\cite{plotkin-swing_hybrid_2020,ohara2023hightemperature}, showing that the intensity of transverse phonon bands is comparable to that of longitudinal ones. 
Investigating this difference between FRFPMS and the new method proposed here, particularly in the context of multiple scattering, is important but requires further study beyond the scope of this manuscript.} 

%
%
%
\begin{figure}
    \centering
    \includegraphics[width=\columnwidth]{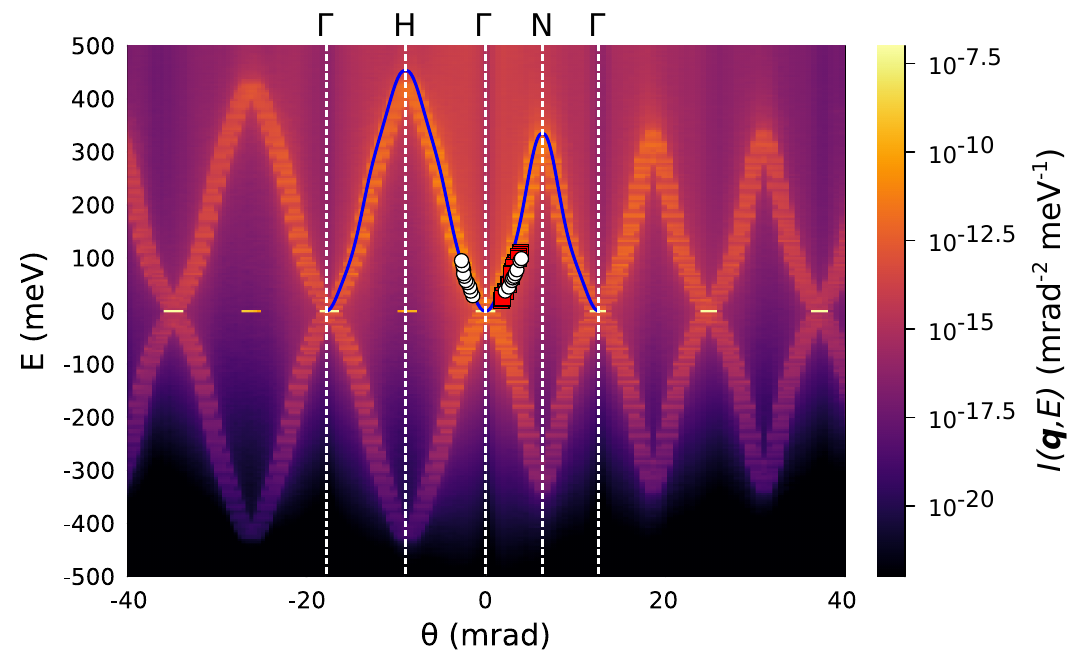}
    \caption{\note{$I(\mathbf{q},E)$ for inelastic magnon scattering on 20~nm thick bcc Fe. The blue curve corresponds to the adiabatic magnon dispersion, while the scattered points are experimental neutron scattering data from Ref. \cite{Mookexperimentaliron1973}. The red squares come from pure Fe, while the white circles correspond to Fe with 4-at.\ \% of Si.}
    }\label{fig:qEplot_magnon}
\end{figure}
%

Finally, we calculate EELS and EEGS due to excitations of magnons. Leveraging the formal analogies between phonons and magnons, we draw parallels as follows: Configurations $\bm{\tau}$ representing atomic displacements are reinterpreted as configurations of tilts of magnetic moments; crystal states $|\mathbf{n}\rangle$ become Fock space vectors of magnon mode occupation numbers, the auxiliary electron beam wave function $\phi(\mathbf{q},\bm{\tau})$ is computed using a magnetic-field-aware method, such as \note{the} Pauli multislice \note{method}~\cite{edstrom_elastic_2016,edstrom_magnetic_2016}, and the MD calculations \note{are} substituted with atomistic spin dynamics calculations~\cite{Eriksson2017}. Furthermore, time scales and energy scales of phonon and magnon dynamics are within the same orders of magnitude.
%
%
These analogies were already exploited in our previous works~\cite{lyon_theory_2021,castellanos_unveiling_2023}, and are extended here to the spectroscopic domain. As a model system for magnon simulations, we selected bcc iron. Following the methodology outlined in Ref.~\cite{castellanos_unveiling_2023}, atomistic spin dynamics simulations \cite{UppASD} were conducted on a $20 \times 20 \times 70$ supercell of bcc iron unit cells with a lattice parameter of 2.87~\AA{} \note{(i.e., with a thickness of 20~nm)} and magnetic moments of 2.30~$\mu_B$ (where $\mu_B$ is the Bohr magneton), computed ab initio along with exchange interactions using the scalar-relativistic \textsc{SPRKKR} code~\cite{Ebert2011}. 
%
%
\footnote{\note{
Employing a time step of 0.1~fs and a Gilbert damping parameter $\alpha=10^{-4}$, a thermalization phase of 100,000 steps was followed by a 0.02~ns trajectory at 300~K for generating snapshots. In total, 5000 snapshots were generated with a $\Delta t=4$~fs, enabling exploration of magnon frequencies up to 125~THz ($\sim 517$~meV). The exit wave functions were computed using the Pauli multislice method on a numerical grid of $1000 \times 1000$ with $2100$ slices across the supercell's thickness. A 200-kV electron probe with a 1~mrad convergence semi-angle, propagating along the [001] direction, was employed. We have used the parametrized magnetic vector potential developed in Ref.~\cite{lyon_parameterization_2021}. The Debye-Waller factor and an absorptive optical potential (see appendix B of Ref.~\cite{castellanos_unveiling_2023}) were included to simulate, in first approximation, the effect of phonon excitations on elastic scattering. For the analysis of $I(\mathbf{q},E)$, we averaged over 48 sets of 250 consecutive snapshots (i.e., $T=1.0$~ps), mutually offset by 100 snapshots.
}}

Figure~\ref{fig:qEplot_magnon} presents a visualization of $I(\mathbf{q},E)$. Similar to the phonon case, the energy gain part of the spectra gradually diminishes with increasing energy transfer $E$, in contrast to the energy loss portion. Also, the characteristic features of magnon band dispersions are revealed by the local maxima of the intensities. 
This is highlighted by the overlap of such maxima with \note{the magnon dispersion at 0~K \footnote{Derived from the adiabatic magnon spectrum~\cite{Eriksson2017} (in the $\Gamma\rightarrow H \rightarrow \Gamma$ path for the horizontal direction, $\theta_y=0$, and in the $\Gamma \rightarrow N \rightarrow\Gamma$ for the $\theta_y=-\theta_x$ diagonal direction).} (blue curve) and the experimental neutron scattering data from Ref. \cite{Mookexperimentaliron1973}.}

In conclusion, we have introduced a method for calculating angle-resolved electron energy loss spectra resulting from phonon or magnon excitations. This approach extends the quantum excitation of phonons model into the energy-loss and energy-gain domains, includes multiple inelastic events, and is straightforward to implement, adding a robust and efficient theoretical tool to the flourishing field of ultra-low energy-loss spectroscopy. 
%

%
%
\begin{acknowledgments}
We acknowledge the support of the Swedish Research Council (grant no.\ 2021-03848), Olle Engkvist's foundation (grant no.\ 214-0331), STINT (grant no. CH2019-8211), and Knut and Alice Wallenbergs' foundation (grant no.\ 2022.0079). The simulations were enabled by resources provided by the National Academic Infrastructure for Supercomputing in Sweden (NAISS) at NSC Centre partially funded by the Swedish Research Council through grant agreement no.\ 2022-06725.
\end{acknowledgments}

%
%


%
%


%

\bibliography{references}


\end{document}